\documentstyle[preprint,prb,aps]{revtex}
\begin{document}
\draft
\begin{title}
{Correlation effects in a quantum dot at high magnetic fields
}
\end{title}
\author{  Kang-Hun  Ahn, J. H. Oh, and K. J. Chang }
\address{ Department of Physics, Korea Advanced Institute
          of Science and Technology, 373-1 Kusung-dong,
          Yusung-ku, Taejon, Korea }
\date{\today}
\maketitle
\begin{abstract}
We investigate the effects of electron correlations on the ground state
energy and the chemical potential of a droplet confined by a parabolic
potential at high magnetic fields.
We demonstrate the importance of correlations in estimating the transition
field at which the first edge reconstruction of the maximum density droplet
occurs in the spin polarized regime.

\end{abstract}
\pacs{  }
\narrowtext

The electronic structure of a quantum dot has been of great interest
both theoretically and experimentally.
The chemical potential $\mu_N$, the energy required to add one electron
to a dot, of a droplet of $N$ electrons can be directly measured from
the conductance peaks resulted from resonant tunneling of an electron
through a quantum dot.\cite{ash,mce}
The magnetic field induces spin and orbital angular momentum changes of
the ground state, thus, the resulting $B-N$ phase diagram has been a subject
of many theoretical studies.\cite{eric,pawel,pal}
The rich dependence of $\mu_N$ on the applied magnetic field provides
information on electron-electron interactions, especially, in the spin
polarized regime, because any transition in this regime results from
many-body phenomena which cannot be explained by a single electron picture.
In recent experiments, Klein and his coworkers \cite{klein} reported a
divergent spin susceptibility, which was explained successfully by
Hartree-Fock (HF) theory.\cite{chamon}
However, the HF theory failed to predict the size of the magnetic field
window in which the maximum density droplet (MDD) state \cite{mac} is the
ground state.

In this paper, we examine the correlation effects in a droplet containig
a small number of electrons and calculate the ground state energy and the
chemical potential via an exact treatment of electron-electron interactions
which includes Landau level mixings.
Our calculations show that the region of magnetic fields in which the ground
state of the spin polarized droplet is the MDD state is much reduced,
compared with the HF calculations, demonstrating the importance of
correlations.
With the Landau level mixing, we also find that magnetizations are reduced and
susceptibility peaks are shifted to higher magnetic fields.
We find bumps in the chemical potentials with varying the magnetic field,
which gives rise to a {\it ripple} in the $\mu-B$ diagram which was observed
by single electron capacitance spectroscopy.\cite{ash}

For simplicity, we first consider the Hamiltonian ($H$) for a spin polarized
droplet of $N$ electrons confined by a parabolic potential
$\frac{1}{2}m^* \omega_0^2 r^2$.
The generalization to an arbitrary spin case is straightforward.
In the strong magnetic limit where all electrons are in the lowest Landau
level,
many-electron states in symmetric gauge with total angular momentum $L$ can be
written as \cite{kinaret}
\begin{eqnarray}
\Psi_{L,N} &  = &  \prod_{i<j} (z_{i}-z_{j})
P_{M} ( z_1,z_2,\ldots,z_{N} )
e^{- \frac{1}{4l^2} \sum_k \mid z_k \mid^2},
\label {general}
\end{eqnarray}
where $z_k = x_k - i y_k$ gives the coordinate of the $k$th electron and
the effective magnetic length $l$ is given by $l^2$ = $\frac{\hbar}{2m^*
\omega_0 (B)}$, where $\omega_0 (B)$ = $\sqrt {\omega_0^2 + \omega_c^2/4}$
and $\omega_c$ is the cyclotron frequency.
$P_{M} ( z_1,z_2,\ldots,z_{N})$ is an $M$th order symmetric polynomial of
N variables, where $M [= L - N(N-1)/2]$ is the excess angular momentum.
For each angular momentum $L$, the energy can be obtained by diagonalizing
the Hamiltonian matrix for all possible $N$-particle basis functions
$\Psi_{L,N}$.
Once the ground state $\mid L,N \rangle$ is obtained, the contribution
from the Coulomb interaction to the ground state energy can be expressed,
\begin{eqnarray}
\langle L,N \mid \sum_{i<j} \frac{e^2}{\epsilon r_{ij} }
\mid L,N \rangle =\frac{e^2}{\epsilon l} C(L,N),
\label{relation}
\end{eqnarray}
where $\epsilon$ is the dielectric constant and $C(L,N)$ is a dimensionless
quantity which only depends on $L$ and $N$.

To study the electronic structure of the droplet, previous theoretical attempts
were mostly relied on Hartree-Fock theory \cite{chamon} or exact calculations
which use a basis set in the Hilbert space restricted to the lowest one or two
Landau levels.\cite{eric,pal}
In our calculations, instead of extending the basis states to those belonging
to higher Landau level indices, which are computationally too expensive,
we employ a symmetric Jastrow correlation factor,\cite{fahy}
which will be multiplied to $\Psi_{L,N}$ so that electron correlations
are effectively included through the mixing of higher Landau levels.
We test various kinds of the Jastrow factor using a variational Monte Carlo
method and find that a simple form of exp$[-\sum_{i<j} u(r_{ij})]$, where
$u(r) = ar + br^2$, gives the lowest energy as shown in Table \ref{tb1}.
Throughout this work, however, we set $a$ = 0 to reduce computational demands
and to perform computations in an analytical fashion.

With information on $C(L,N)$ in Eq. (\ref{relation}), the ground state energy
in the variational many-body state, $\Phi_{L,N}(b)$ = $e^{-b \sum_{i<j}
r_{ij}^2} \Psi_{L,N}$, can be evaluated in an analytical fashion.
First, let us decompose the Hamiltonian into two parts, $H_{\Omega}$ and
$V_{\Omega}$,
where
\begin{eqnarray}
H_{\Omega}  &  =  &  \sum_{i} [
  \frac{(\vec{p}_i -\frac{e}{c}\vec{A})^2}{2 m^*}
  + \frac{1}{2}m^* \omega_0^2r_i^2]
  -  \frac{1}{2} g^* \frac{m^*}{m_0} \hbar \omega_c S_{z}
  - \sum_{i<j} \frac{1}{2}m^* \Omega^2 r_{ij}^2,
\label{homega}
\end{eqnarray}
\begin{eqnarray}
V_{\Omega}
   & = &  \sum_{i<j}  \Bigl(\frac{1}{2}m^*  \Omega^2 r_{ij}^2 + \frac{e^2}
  {\epsilon \mid \vec{r}_i -\vec{r}_j \mid} \Bigr),
\label{vomega}
\end{eqnarray}
where the total spin angular momentum $S_z$ is $N$/2 for a spin polarized
droplet.
If the parameter $b$ in $\Phi_{L,N}(b)$ is replaced by $m^* \frac{\Omega_0
-\omega_0 (B)} {2 N \hbar}$, where $\Omega_0 = \sqrt {\omega_0 ^2 (B) - N
\Omega^2}$, $\Phi_{L,N}(\Omega)$ will be the eigenstates of
$H_{\Omega}$.\cite{johnson}
Then, the eigenvalues of $H_{\Omega}$ for the states of zero center-of-mass
angular momentum (a general case will be discussed later) are
\begin{eqnarray}
H_{\Omega} \Phi_{L,N} (\Omega)
& = &
\left[ E_{\Omega}(N) + L
\hbar(\Omega_0-\omega_c/2) \right] \Phi_{L,N} (\Omega),
\label{harmo}
\end{eqnarray}
where
\begin{eqnarray}
E_{\Omega}(N) & = & \hbar \omega_0(B) + (N-1) \hbar \Omega_0
- \frac{1}{2} g^* \frac{m^*}{m_0} \hbar \omega_c S_{z}.
\label{energ}
\end{eqnarray}
The expectation value of the first term (an {\it ad hoc} harmonic potential)
in Eq. (\ref{vomega}) can be evaluated analytically by using the raising
($a_{ij}^{+}$ and $b_{ij}^{+}$) and lowering ($a_{ij}^{-}$ and $b_{ij}^{-}$)
operators defined as,\cite{johnson}
\begin{eqnarray}
a_{ij}^{+} & = & \left[ \frac{1}{4m^* \hbar \Omega_0}\right]^{\frac{1}{2}}
\left[ m^*\Omega_0 z_{ij}^{~} -i (p_{ij,x} - i p_{ij,y})\right],
\label {op1}
\end{eqnarray}
\begin{eqnarray}
b_{ij}^{+} & = & \left[ \frac{1}{4m^* \hbar \Omega_0}\right]^{\frac{1}{2}}
\left[ m^*\Omega_0 z_{ij}^* -i (p_{ij,x} + i p_{ij,y})\right],
\label {op2}
\end{eqnarray}
where $a_{ij}^{-}$($b_{ij}^{-}$) is the hermitian conjugate of
$a_{ij}^{+}$($b_{ij}^{+}$), $\vec{p}_{ij} = \vec{p}_i - \vec{p}_j =
(p_{ij,x},p_{ij,y})$, and $z_{ij} = z_i - z_j$.
Then, it follows that
\begin{eqnarray}
\frac{m^*\Omega_0}{\hbar} r_{ij}^2
= a_{ij}^{+} a_{ij}^- + b_{ij}^{+} b_{ij}^- + a_{ij}^{+} b_{ij}^+
+ a_{ij}^{-} b_{ij}^- + 2.
\end{eqnarray}
Considering the states of zero center-of-mass angular momentum, the polynomial
$P_M$ in Eq. (\ref{general}) can be expressed only in terms of relative
coordinates, and the variational many-body state is rewritten with the use of
$z_{ij}=(\frac{\hbar}{m^* \Omega_0})^{1/2}(a_{ij}^{+}+b_{ij}^{-})$,
\begin{eqnarray}
\Phi_{L,N} (\Omega) & = & \prod_{i<j} z_{ij} P_{M}(\sum_{i \neq 1} z_{1i},
\sum_{i \neq 2} z_{2i} , \ldots, \sum_{i \neq N } z_{Ni} ) \Psi_v \\
   & = & (\frac{ \hbar}{m^* \Omega_0})^{L/2}
   \prod_{i<j} a_{ij}^{+} P_{M}(\sum_{i \neq 1} a_{1i}^{+},
   \sum_{i \neq 2} a_{2i}^{+} , \ldots , \sum_{i \neq N} a_{Ni}^{+} ) \Psi_v,
\end{eqnarray}
where $\Psi_v$ denotes the vacuum state, {\it i.e.}, $a_{ij}^{-} \Psi_v =
b_{ij}^{-} \Psi_v = 0$, and is given by $\Psi_v$ = exp$[ -\frac{N m^*
\omega_0 (B)}{2 \hbar} R^2 - \frac{m^* \Omega_0}{2 N \hbar} \sum_{i<j}
r_{ij}^2 ]$ where $\vec{R}$ is the center-of-mass coordinate.
Using the commutation relations of $a_{ij}^{\pm}$ and $b_{ij}^{\pm}$,
the expectation value of the {\it ad hoc} harmonic potential is
\begin{eqnarray}
\left\langle \Phi_{L,N}(\Omega)
\left  | \frac{1}{2} m^* \Omega^2 \sum_{i<j} r_{ij}^2
\right | \Phi_{L,N}(\Omega) \right\rangle
&\! =\! &
\frac{\hbar \Omega^2}{2 \Omega_0}
\left \langle
\Phi_{L,N}(\Omega)
\left  |   \sum_{i<j} ( a_{ij}^+ a_{ij}^- + 2)
\right |  \Phi_{L,N}(\Omega)
\right \rangle
\nonumber \\
&\! =\! &
\frac{\hbar \Omega^2}{2 \Omega_0}
[  N L + N(N-1) ].
\label{har}
\end{eqnarray}
For the Coulomb interaction energy in the variational state $\Phi_{L,N}
(\Omega)$, replacing $l$ by $\Gamma_{\Omega} (= \sqrt {\frac{\hbar}{2m^*
\Omega_0}})$
in Eq. (\ref{relation}) leads to the expectation value,
\begin{eqnarray}
\langle \Phi_{L,N}(\Omega) \mid \sum_{i<j} \frac{e^2}{\epsilon r_{ij} }
\mid \Phi_{L,N}(\Omega) \rangle =\frac{e^2}{\epsilon \Gamma_{\Omega}} C(L,N).
\label{genrel}
\end{eqnarray}

{}From Eqs. (\ref{harmo}), (\ref{har}), and (\ref{genrel}), we finally get
the total energy $E_{L,N}$ for total angular momentum $L$ and zero
center-of-mass angular momentum by minimizing $\langle \Phi_{L,N} (\Omega)
\mid H \mid \Phi_{L,N} (\Omega) \rangle$ over the variable $\Omega$;
\begin{eqnarray}
E_{L,N} & = & min_{\Omega} \{ E_{\Omega}(N) + L
\hbar (\Omega_0-\omega_c/2)
+\frac{\hbar \Omega^2}{2 \Omega_0} [ NL + N(N-1) ]
+\frac{e^2}{\epsilon \Gamma_{\Omega}} C(L,N) \}.
\label {flp}
\end{eqnarray}
In general, the total angular momentum is given by $L = M_r + M_c$,
where $M_r$ and $M_c$ represent the angular momentums of the
relative and center-of-mass motions, respectively.
If $M_c \neq 0$, the total energy should include an additional energy,
$M_c \hbar [ \omega_0 (B) -\omega_c/2 ]$, with $L$ replaced by $M_r$ in
Eq. (\ref{flp}).
However, one can see that the states with $M_c \neq 0$ give higher
energies for given $N$ and $B$.
Thus, the ground state energy of the $N$-electron droplet at magnetic field
$B$ is a minimum value of $E_{L,N}$ calculated for $L \ge \frac{N(N-1)}{2}$
and $M_c$ = 0.

For a droplet with arbitary spin states, we can generalize the total energy
$E_{L,N}$ to $E_{L,S_{z},N}$ by replacing $C(L,N)$ with $C(L,S_{z},N)$,
which is defined analogously to Eq. (\ref{relation}).
In this case, instead of $\Psi_{L,N}$ in Eq. (\ref{general}), we use
$\Psi_{L,S_{z},N}$ defined as,
\begin{eqnarray}
\Psi_{L,S_{z},N}= e^{- \frac{1}{4l^2} \sum_k \mid z_k \mid^2}
\sum_{n} Q^{(n)}_{L}(z_{1},z_{2},\ldots ,z_{N}) \chi_{n},
\label{lsn}
\end{eqnarray}
where $Q^{(n)}_{L}$ is an $L$th order polynomial of $N$ variables, $\chi_n$
is a $N$-product of single particle spinors, and $n$ denotes possible
spinor configurations.
Although $\Psi_{L,S_{z},N}$ is totally antisymmetric, each of $Q^{(n)}_{L}$
and $\chi_{n}$ does not necessary satisfy the antisymmetric property.

The ground state energies in the variational many-body state $\Phi_{L,S_z,N}$
for a six-electron quantum dot are plotted as a function of magnetic field
in Fig. \ref{fig1}.
Throughout this work, we use the confinement strength of $\hbar \omega_0$
= 2.5 meV and the usual values for $\epsilon$ and $m^*$ in GaAs.
Compared with the results of conventional exact calculations using
$\Psi_{L,S_z,N}$ in Eq. (\ref{lsn}), our calculated energies are
generally lower with significant reductions at lower magnetic fields.
This is because the Jastrow factor introduces more effectively the Landau
level mixing and the correlation effect in the region of low magnetic fields,
which are absent in the conventional scheme.
{}From Fig. \ref{fig1}, one can notice that with the Jastrow factor
magnetizations $(- dE/dB)$ are relatively decreased whereas susceptibility
peaks are shifted to higher magnetic fields.
We also note that the variational ground states exhibit different orbital
and spin angular momentum quantum numbers, compared with the conventional
exact calculations.

At low magnetic fields, 0 $< B < B_c$ = 1.1 T, the quantum dot is found to
favor the equal occupations of the up- and down-spin states, {\it i.e.,}
$L$ = 6 and $S_z$ = 0, while the minimum total spin state only appears
in a very narrow range near zero field in the conventional calculations
without the Jastrow factor.
At $B_c$, the transfer of electrons between the up- and down-spin states
begins and continues due to the Zeeman splitting energy up to the value of
$B_I$ = 1.25 T, where a spin polarized state appears.
With varying the magnetic field, the bumps in the chemical potentials
($\mu_N$ = $E_N$ - $E_{N-1}$), which are associated with the ground state
transitions, are clearly seen in Fig. \ref{fig2}, exhibiting a ripple in
the $\mu_{N}-B$ diagram.
In fact, such a feature in the $\mu_{N}-B$ graph was observed in
single electron capacitance spectroscopy measurements.\cite{ash}

Our calculations show a spin polarized droplet ($S_z$ = $N$/2) above
$B_I$, and a magnetic field window in which the ground state of
the spin polarized droplet is the maximum density droplet [$L$ = $N(N-1)$/2]
where electrons occupy one-electron eigenstates of angular momentum indices,
$m$ = 0, 1, ..., and ($N - 1$).
As the magnetic field increases, since the Coulomb interaction energy
grows due to the reduction of the radius of the MDD, the MDD must undergo
a transition to a higher angular momentum state ($L$ = 20 and $S_z$ = 2)
to lower the ground state energy.
In the $N$ = 6 droplet, the edge of the MDD is reconstructed at
$B_r$ = 3.7 T, and at this field an abrupt upward shift of the chemical
potential appears, as shown in Fig. \ref{fig2}.
Such a step at $B_r$ is the reminiscent of the chemical potential jump in the
fractional quantum Hall effect, which is caused by quasi-hole
creation.\cite{fqhe}
Indeed, experimentally, an upward step in $\mu_{N}$ at $B_r$ was observed
for a droplet in the spin polarized regime.\cite{klein}

Recently, Klein and his coworkers observed many transition fields in a droplet
of $N \sim 30$ electrons, and showed that the HF theory describes successfully
their experimental data with a fitting parameter $\hbar \omega_0$
for $B \leq B_{I}$.\cite{klein}
However, the transition field $B_r$ was overestimated severely.
Because of the computational difficulties for higher values of $N$,
our calculations are restricted to droplets of $N \leq 7$.
Nevertheless, we are able to see the effect of electron correlations on
the transition fields.
By varying the confinement strength from 1 to 5 meV for the $N$ = 7 droplet,
we plot the size of the magnetic field window, ($B_r - B_I$), in which
the MDD is the ground state of the spin polarized droplet, as a function of
$B_I$ in Fig. \ref{fig3}.
With the Jastrow correlation factor, we find that the size of ($B_r - B_I$)
is significantly reduced by about 1 T, compared with the HF calculations,
while in the conventional exact calculations with only the lowest Landau level
the reduction is about 0.5 T.
Thus, it is important to include the correlation effect in predicting the
transition fields for the MDD state.

In conclusion, we have made a detailed study of the correlation effect
on the ground state, the chemical potential, and the transition fields
in strong magnetic fields.
Using an exact treatment of electron-electron interactions through the
Landau level mixing, we find that magnetizations are reduced while the
transition fields are shifted to higher magnetic fields by the correlation
effect.
We present a ripple in the $\mu_N - B$ diagram, which has not been explained
by previous calculations.
In estimating the size of the magnetic field window in which the MDD appears,
we attribute the failure of the HF theory to the neglect of electron
correlations.

\acknowledgments

This work was supported by the CMS at KAIST.

\begin{figure}
\caption{ The ground state energies vs magnetic field for a six-electron
quantum dot with the confinement strength of $\hbar \omega_0$ = 2.5 meV.
The arrows indicate the transition fields at which the ground state crossings
occur.
Some chosen ground states are labeled by the quantum numbers ($L, 2S_z$).}
\label{fig1}
\end{figure}

\begin{figure}
\caption{ (a) The chemical potential $\mu_N$ vs magnetic field for
$\hbar \omega_0 = 2.5$ meV.
(b) The details of $\mu_N$ for $N$ = 7.
Curve segments between two downward arrows are labeled by the ground state
quantum numbers ($L, 2S_z$) of the $N$ = 7 droplet, while those between two
upward arrows correspond to the ground states of the $N$ = 6 droplet.
}
\label{fig2}
\end{figure}

\begin{figure}
\caption{For the spin polarized droplet of $N$ = 7 electrons, ($B_r - B_I$)
is plotted as a function of $B_I$ which is determined by varying the
confinement strength from 1 to 5 meV.
The present results with the Jastrow correlation factor are compared
with the Hartree-Fock and conventional exact calculations without the
Jastrow factor.
 }
\label{fig3}
\end{figure}

\begin{table}
\caption{ The variational ground state energies calculated by a Monte Carlo
method are compared for various forms of the Jastrow factor, $\exp[-\sum_
{i<j}u(r_{ij})]$. We use $N$ = 3, $B$ = 3 T, and $\hbar \omega_0$ = 5.4 meV.}
\label{tb1}
\begin{tabular}{cccccc}
$u(r)$              &  Ground state energy (meV)  \\
     \hline
$ br^2 $            & 41.041  \\
$ ar + br^2 $       & 40.866 $\pm$ 0.005 &  \\
$ a \log(r)$        & 41.248 $\pm$ 0.010 &   \\
$br/(1+ar)$         & 41.230 $\pm$ 0.010 &   \\
$ \frac{a}{\sqrt{r}}[ 1 - \exp( -\sqrt{r/b}-r/2b) ] $ & 41.194 $\pm$ 0.005 &
\\
$ a[1-\exp(-br)]/r$  & 41.750 $\pm$ 0.100 &   \\
\end{tabular}
\end{table}

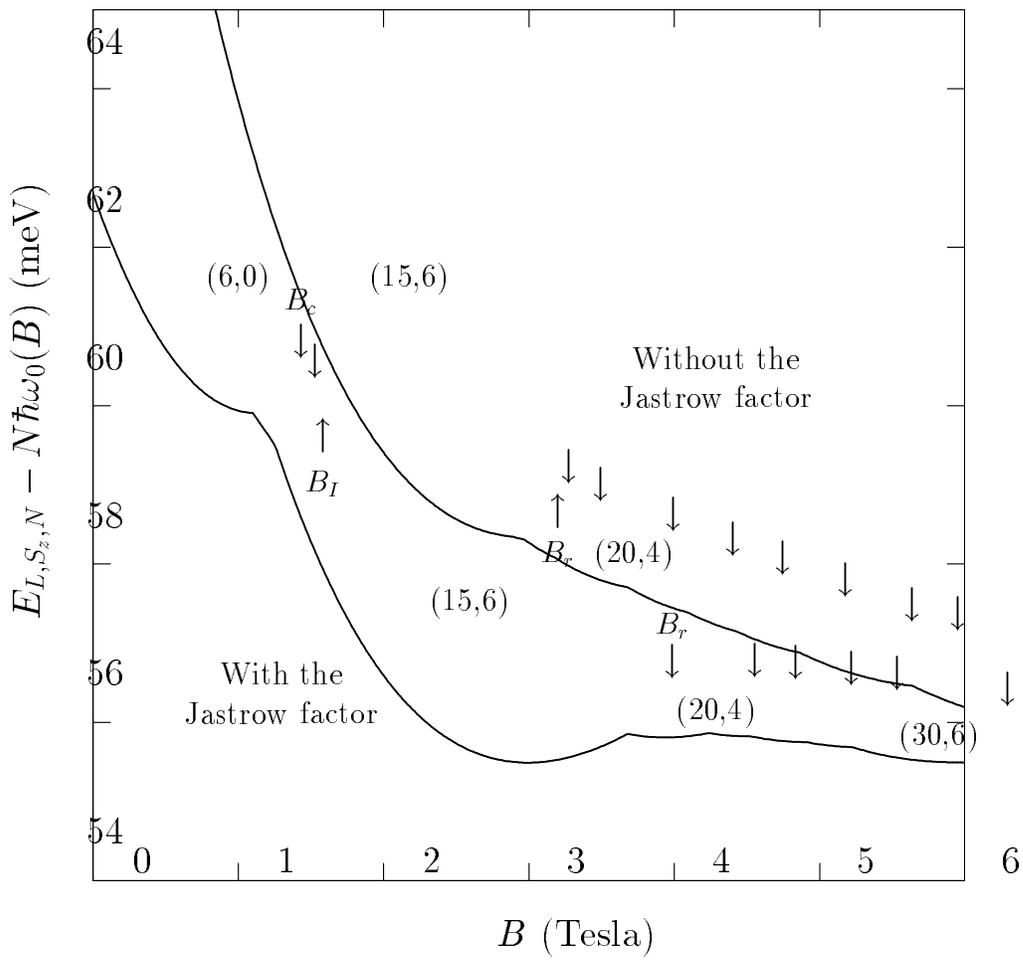
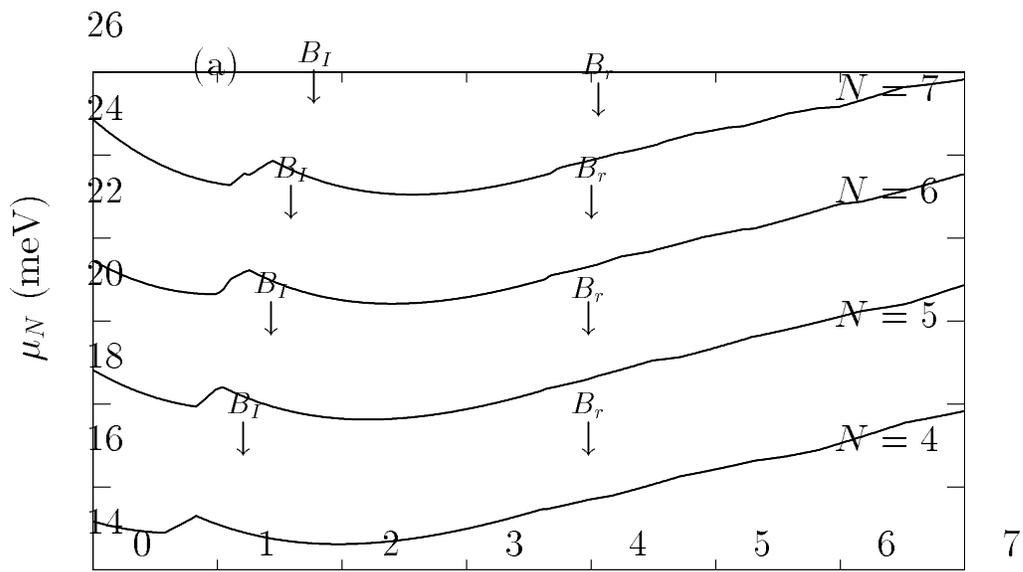
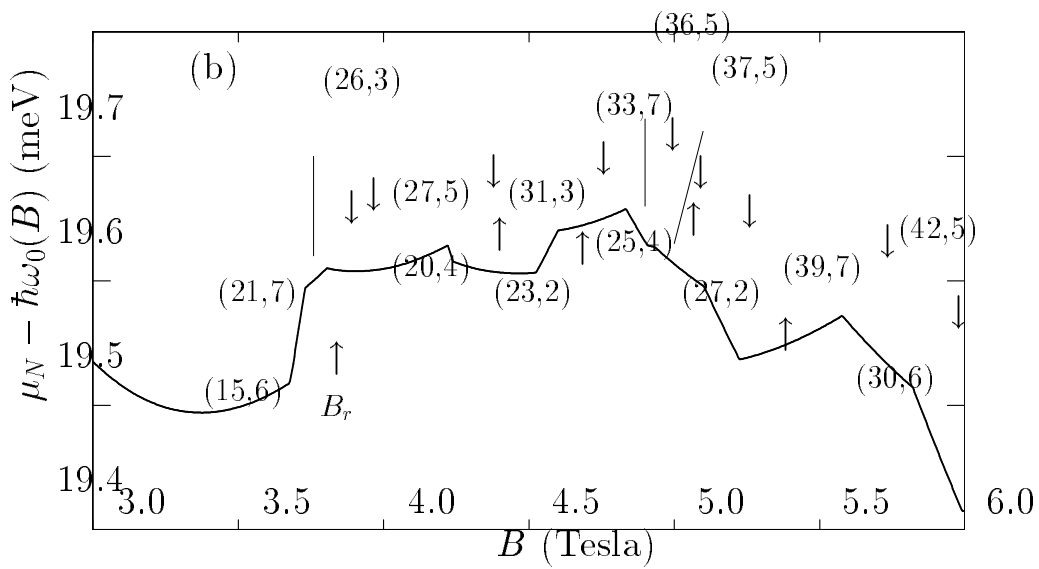
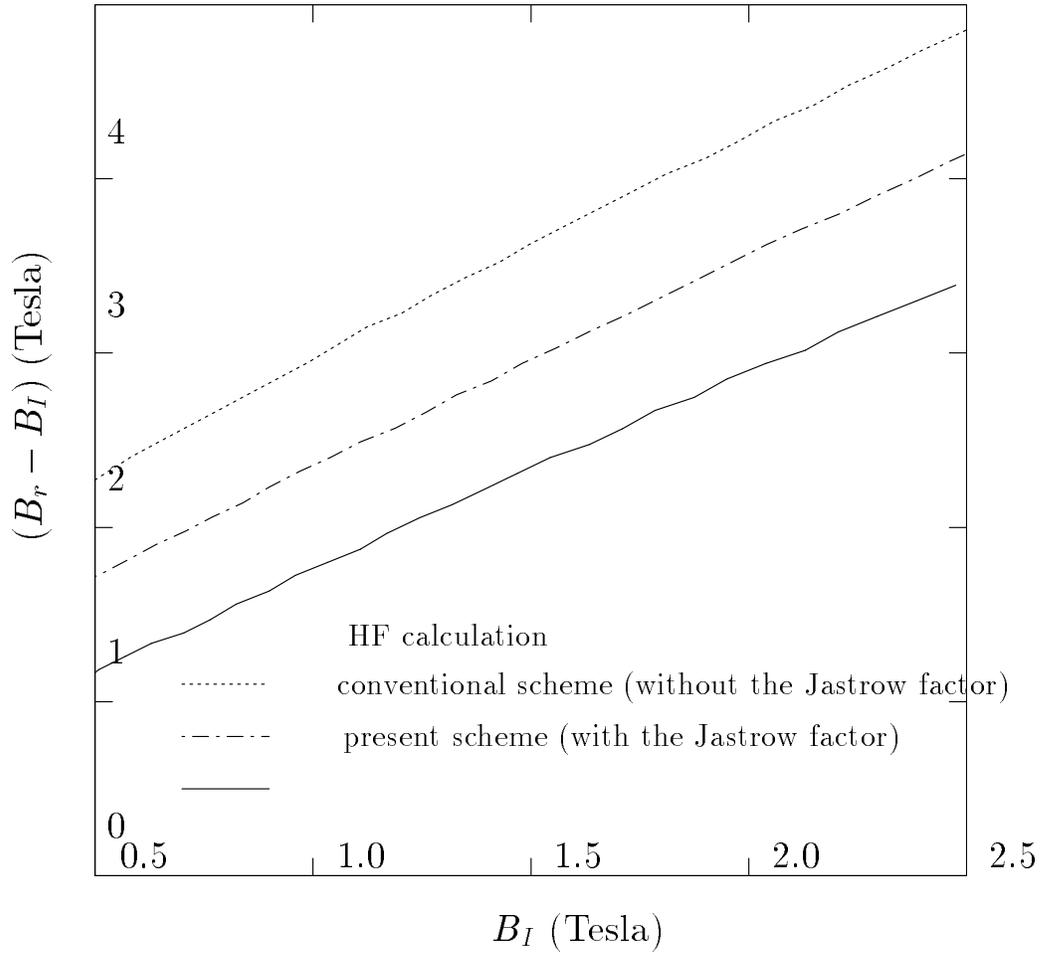
\end{document}